%% file: pbpbjetmain.tex
\newif\ifdraft
\newif\iffull
\newif\ifcomment
\newif\iflatexdiff
\newif\ifbibtex
\newif\ifpreprint
\newif\ifraatwopanels

\latexdifftrue    
\fulltrue        
\bibtextrue       
\preprinttrue     
\raatwopanelstrue 

\def\dvers{v2.1}
\def\snntitle{$\snn$}
\ifpreprint
\def\snntitle{$\snnbf$}
\fi
\def\dtitle{Measurement of jet suppression \\ in central Pb--Pb collisions at \snntitle = 2.76 TeV} 
\def\stitle{Measurement of jet suppression in central Pb--Pb collisions} 
\ifpreprint
\documentclass[ALICE,manyauthors,12pt]{cernphprep}
\usepackage[comma,square,numbers,sort&compress]{natbib}
\else
\documentclass[final,3p,12pt]{elsarticle}
\biboptions{comma,square,numbers,sort&compress}
\fi
\usepackage{hyperref}
\usepackage{graphicx}  
\usepackage{dcolumn}   
\usepackage{bm}        
\usepackage{amssymb}   
\usepackage{amsfonts}
\usepackage{graphics}
\usepackage{grffile}   
\usepackage{epsfig}
\usepackage{units}
\usepackage{comment}
\usepackage[usenames]{color}
\usepackage[normalem]{ulem} 
\usepackage{color}
\usepackage[utf8]{inputenc}
\usepackage[T1]{fontenc}
\usepackage{subfigure}
\iflatexdiff
\RequirePackage{color}\definecolor{RED}{rgb}{1,0,0}\definecolor{BLUE}{rgb}{0,0,1}

\fi
\ifpreprint
\else
 
\fi
\input{commands.tex}
\graphicspath{{./img/}}
\ifdraft
\usepackage{lineno}
\linenumbers
\modulolinenumbers[5]
\usepackage{fancyhdr}
\else
\renewcommand{\warn}[1]{}

\fi

\begin{document}
\newlength{\figlen}
\setlength{\figlen}{\linewidth}
\ifpreprint
\setlength{\figlen}{0.95\linewidth}
\begin{titlepage}
\PHyear{2015}
\PHnumber{023}                  
\PHdate{30 January}             
\title{\dtitle}
\ShortTitle{\stitle}
\Collaboration{ALICE Collaboration%
         \thanks{See Appendix~\ref{app:collab} for the list of collaboration members}}
\ShortAuthor{ALICE Collaboration} 
\ifdraft
\begin{center}
\today\\ \color{red}DRAFT \dvers\ \hspace{0.3cm} \$Revision: 2031 $\color{white}:$\$\color{black}\vspace{0.3cm}
\end{center}
\fi
\else
\begin{frontmatter}
\title{\dtitle}
\iffull
\input{authors-plb.tex}
\else
\ifdraft
\author{ALICE Collaboration \\ \vspace{0.3cm} 
\today\\ \color{red}DRAFT \dvers\ \hspace{0.3cm} \$Revision: 2031 $\color{white}:$\$\color{black}}
\else
\author{ALICE Collaboration}
\fi
\fi
\fi
\begin{abstract}
The transverse momentum~(\pT) spectrum and nuclear modification factor~(\RAA) of reconstructed jets 
in $0$--$10$\% and $10$--$30$\% central \PbPb\ collisions at $\snn=2.76$~TeV were measured.
Jets were reconstructed using the anti-\kT~jet algorithm with a resolution parameter of $R=0.2$ from charged and neutral particles, 
utilizing the \mbox{ALICE} tracking detectors and Electromagnetic Calorimeter (EMCal).
The jet $\pt$ spectra are reported in the pseudorapidity interval of $|{\eta}_{\rm jet}|<0.5$ 
for $40<\jetpt<120$~\GeVc\ in $0$--$10$\% and for $30<\jetpt<100$~\GeVc\ in $10$--$30$\% collisions.
Reconstructed jets were required to contain a leading charged particle with $\pT>5$~\GeVc\ to suppress
jets constructed from the combinatorial background in \PbPb\ collisions.
The leading charged particle requirement applied to jet spectra both in \pp\ and \PbPb\ collisions 
had a negligible effect on the \RAA. 
The nuclear modification factor 
\RAA\ was found to be $0.28\pm0.04$ in $0$--$10$\% and $0.35\pm0.04$ in \mbox{$10$--$30$\%} collisions, 
independent of $\jetpt$ within the uncertainties of the measurement.  
The observed suppression is in fair agreement with expectations 
from two model calculations with different approaches to jet quenching.

\ifdraft 
\ifpreprint
\end{abstract}
\end{titlepage}
\setcounter{page}{2}
\else
\end{abstract}
\end{frontmatter}
\newpage
\fi
\fi
\ifdraft
\thispagestyle{fancyplain}
\else
\end{abstract}
\ifpreprint
\end{titlepage}
\setcounter{page}{2}
\else
\end{frontmatter}
\fi
\fi
\input{pbpbjetcontent.tex}

\ifpreprint
\iffull
\newenvironment{acknowledgement}{\relax}{\relax}
\begin{acknowledgement}
\section*{Acknowledgements}
\input{acknowledgements_jan2015.tex}        
\end{acknowledgement}
\ifbibtex
\bibliographystyle{utphys}
\bibliography{biblio}{}
\else
\input{refpreprint.tex}
\fi
\newpage
\appendix
\section{The ALICE Collaboration}
\label{app:collab}
\input{Alice_Authorlist_2015-Jan-29.tex}      
\else
\ifbibtex
\bibliographystyle{utphys}
\bibliography{biblio}{}
\else
\input{refpreprint.tex}
\fi
\fi
\else
\iffull
\input{acknowledgements_jan2015.tex}
\input{refpaper.tex}
\else
\ifbibtex
\bibliographystyle{utphys}
\bibliography{biblio}{}
\else
\input{refpaper.tex}
\fi
\fi
\fi
\end{document}

%% file: commands.tex
\newcommand{\com}[1]       {\relax}

\newcommand{\pp}           {pp}

\newcommand{\PbPb}         {\mbox{Pb--Pb}}

\newcommand{\s}            {\ensuremath{\sqrt{s}}}
\newcommand{\pt}           {\ensuremath{p_{\mathrm{T}}}{ }}

\newcommand{\snn}          {\ensuremath{\sqrt{s_{\mathrm{NN}}}}}
\newcommand{\snnbf}        {\ensuremath{\mathbf{{\sqrt{\textit{s}_{NN}}}}}}

\newcommand{\Ncoll}        {\ensuremath{N_\mathrm{coll}}}

\newcommand{\abs}[1]       {\ensuremath{\left|#1\right|}}

\newcommand{\Dphi}         {\ensuremath{\Delta\varphi}}
\newcommand{\Deta}         {\ensuremath{\Delta\eta}}

\newcommand{\Fig}[1]       {Fig.~\ref{#1}}

\newcommand{\Ref}[1]       {Ref.~\cite{#1}}
\newcommand{\Refs}[1]      {Refs.~\cite{#1}}

\newcommand{\red}[1]       {\textcolor{red}{#1}}

\newcommand{\warn}[1]      {{\small\textbf{\red{(!}\footnote{\textbf{\red{(!)}}~#1}\red{)}}}\marginpar{\textbf{\red{---}}}}

\iflatexdiff

\renewcommand{\xout}[1]    {\textcolor{red}{\sout{#1}}}
 
\else

\renewcommand{\xout}[1]    {}
\fi
\newcommand{\jetpt}        {\ensuremath{{p}_{\mathrm{T,~jet}}}}  
\newcommand{\jeteta}       {\ensuremath{{\eta}_{\mathrm{jet}}}}  
\newcommand{\jetphi}       {\ensuremath{{\varphi}_{\mathrm{jet}}}}  
\newcommand{\dpT}          {\ensuremath{\delta p_{\mathrm{T}}}}  
\newcommand{\pT}           {\ensuremath{p_{\mathrm{T}}}}  
  
\newcommand{\ET}           {\ensuremath{E_{\mathrm{T}}}}  
\newcommand{\rhochem}      {\ensuremath{{\rho}_{\mathrm{scaled}}}}
\newcommand{\rhoch}        {\ensuremath{{\rho}_{\mathrm{ch}}}}
\newcommand{\RAA}          {\ensuremath{{R}_{\mathrm{AA}}}}
\newcommand{\TAA}          {\ensuremath{{T}_{\mathrm{AA}}}}

\newcommand{\FastJet}      {\ensuremath{\mathrm{FastJet}}}
\newcommand{\kT}           {\ensuremath{k_{\mathrm{T}}}}
\newcommand{\jetraw}       {\ensuremath{{p}_{\mathrm{T,\,jet}}^{\mathrm{raw}}}}
\newcommand{\jetreco}      {\ensuremath{{p}_{\mathrm{T,\,jet}}^{\mathrm{reco}}}}
\newcommand{\jettrue}      {\ensuremath{{p}_{\mathrm{T,\,jet}}^{\mathrm{true}}}}
\newcommand{\GeVc}         {GeV/$c$}

\newcommand{\MeVc}         {MeV/$c$}
\newcommand{\Ajet}         {\ensuremath{{A}_{\mathrm{jet}}}}
\newcommand{\qhat}         {\ensuremath{\hat{q}}}

%% file: pbpbjetcontent.tex

\section{Introduction}
\label{sec:intro}
Discrete formulations of Quantum Chromodynamics~(QCD) predict the existence of a cross-over transition 
from normal nuclear matter to a new state of matter called the Quark-Gluon Plasma~(QGP), where the partonic 
constituents, quarks and gluons, are deconfined.
The QGP state is expected to exist at energy densities above $0.5$~GeV/fm$^3$ and temperatures above $160$ MeV~\cite{Bhattacharya:2014ara},
which can be reached in collisions of heavy-ions at ultra-relativistic energies.
The existence of the QGP is supported by the observations reported by experiments at
the Relativistic Heavy Ion Collider~(RHIC)~\cite{Arsene20051,Back200528,Adcox2005184,Adams2005102} and at the Large Hadron 
Collider~(LHC)~\cite{Aamodt:2010pb,Aamodt:2010cz,Chatrchyan:2011pb,Aamodt:2011mr,Aamodt:2010pa,ATLAS:2011ah,Chatrchyan:2012wg,ALICE:2011ab,
Aad:2013xma,Chatrchyan:2013kba,Aamodt:2010jd,Chatrchyan:2011sx}.

One way to characterize the properties of the QGP is to use partons from the hard scattering 
of the partonic constituents in the colliding nucleons as medium probes. 
Hard scattering is expected to occur early in the collision evolution, producing high transverse momentum~(\pT) partons,
which propagate through the expanding medium and eventually fragment into jets of hadrons. 

Due to interactions of the high-\pT\ partons with the medium, the energy of the partons is reduced compared to proton--proton (\pp) 
collisions due to medium-induced gluon radiation and collisional energy loss~(jet quenching)~\cite{Gyulassy:1990ye,Baier:1994bd}. 
The production cross section of the initial hard scattered partons is calculable using perturbative QCD (pQCD), 
and the contribution from the non-perturbative hadronization can be well calibrated via jet measurements in \pp\ collisions. 

Jet quenching has been observed at RHIC~\cite{Adcox:2001jp,Adler:2002tq,Adcox:2002pe,Adler:2003qi,Adams:2003kv,
Adams:2003im,Back:2003qr,Arsene:2003yk,Adare:2006nr,Adare:2008cg} and at the LHC~\cite{Aamodt:2010jd,Aad:2010bu,Chatrchyan:2011sx,Chatrchyan:2011pb,
Aamodt:2011vg,CMS:2012aa,Chatrchyan:2012nia,Chatrchyan:2012gw,Chatrchyan:2012gt,Aad:2012vca,Chatrchyan:2013exa,Chatrchyan:2013kwa,
Chatrchyan:2014ava,Aad:2014wha,Aad:2014bxa}
via the measurement of inclusive hadron and jet production at high $\pT$, di-hadron angular correlations and the dijet energy imbalance. 
In all cases, the measured observable is found to be strongly modified in central heavy-ion collisions relative to \pp\ collisions, 
when compared to expectations based on treating heavy-ion collisions as an incoherent superposition of independent nucleon--nucleon collisions.

Measurements of the jet kinematics are expected to be more closely correlated to the initial parton kinematics than measurements of 
high-\pT~hadrons.
Jets are usually reconstructed by grouping measured particles within a given distance, e.g.\ a cone with radius $R$. 
The interaction with the medium can result in a broadening of the jet shape, a softening of the jet fragmentation~\cite{Salgado:2003rv} 
leading to an increase of out-of-cone gluon radiation~\cite{Vitev:2005yg} with respect to jets reconstructed in \pp\ collisions\cite{Chatrchyan:2011sx}.
Therefore, for a given jet resolution parameter $R$ and a fixed initial parton energy, the energy of jets reconstructed in heavy-ion collisions 
is expected to be smaller than those reconstructed in \pp\ collisions.

Jet measurements in heavy-ion collisions are challenging since a single event can have multiple, possibly overlapping, jets
from independent nucleon--nucleon scatters, as well as combinatoric ``jets'' from the large, partially correlated and fluctuating background of low 
transverse momentum particles.
Consequently, jet reconstruction in heavy-ion collisions requires a robust jet-signal definition,
and a procedure to correct for the presence of the large background and its associated region-to-region fluctuations~\cite{Cacciari:2010te}.

The results reported in this letter are from lead--lead~(\PbPb) collision data at an energy per nucleon pair of $\snn=2.76$~TeV 
recorded by the ALICE detector in 2011.
Charged particles are reconstructed with the Inner Tracking System~(ITS) and the Time Projection Chamber~(TPC) down to \pT\ of $0.15$~\GeVc.
Neutral particles, excluding neutrons and $K^0_{\rm L}$s, are reconstructed with the Electromagnetic Calorimeter~(EMCal) down to a transverse energy 
of the EMCal clusters of $0.3$~GeV.  
For jet reconstruction, we followed the approach applied in \Refs{Abelev:2012ej,Abelev:2013kqa}, where the average energy density of the event was subtracted 
from the signal jets on a jet-by-jet basis, and the detector and background effects were corrected on an ensemble basis via an unfolding procedure.
The signal jets were obtained using the anti-\kT~jet algorithm~\cite{Cacciari:2008gp} 
with a resolution parameter of $R=0.2$ in the pseudorapidity range of $|\jeteta|<0.5$.
Signal jets were required to contain at least one charged particle with $\pT>5$~\GeVc.
The corrected jet $\pt$ spectra and nuclear modification factors ($\RAA$) are reported for $40<\jetpt<120$~\GeVc\ 
in $0$--$10$\% and for $30<\jetpt<100$~\GeVc\ in $10$--$30$\% central \PbPb\ collisions 
and the corrected jet $\pt$ spectrum for $20<\jetpt<120$~\GeVc\ in \pp\ collisions at $\sqrt{s}=2.76$~TeV from 13.6 n${\mathrm{b}}^{-1}$ recorded in 2011.
The \RAA\ is compared to expectations from two jet quenching model calculations with different approaches, described later, in order 
to test the sensitivity of the observable to the energy density via the centrality dependence, and to the parton energy scale via the momentum dependence.

\section{Experimental setup}
\label{sec:setup}
For a complete description of the ALICE detector and its performance see Refs.~\cite{Aamodt:2008zz} and~\cite{Abelev:2014ffa}, respectively.
The analysis presented here relies mainly on the ALICE tracking system and EMCal, 
both of which are located inside a large solenoidal magnet with field strength $0.5$~T. 

The tracking system consists of the ITS, a high-precision six-layer silicon detector system with the inner layer at $3.9$~cm and the outer at $43$~cm
from the center of the detector, and the TPC with a radial extent of $85$--$247$~cm, provides up to 159 independent space points per track.
The two innermost layers of the ITS consist of the Silicon Pixel Detector (SPD), which provides two layers of silicon pixel sensors 
at radii $3.9$~cm and $7.6$~cm from the beam axis and covers the full azimuth over $|\eta|<2$ and $|\eta|<1.4$, respectively.
The combined information of the ITS and TPC can determine the momenta of charged particles from low momentum ($\pT\approx0.15$~\GeVc) 
to high momentum ($\pT\approx100$~\GeVc) in $|\eta|<0.9$ and full azimuth.

The EMCal is a Pb-scintillator sampling calorimeter, which covers $107$ degrees in azimuth and $|\eta|<0.7$. 
It consists of $10$ supermodules with a total of $11520$ individual towers each covering an angular region $\Deta\times\Dphi=0.014\times0.014$ which are read out by avalanche photodiodes. 

The data were recorded in 2011 for \PbPb\ collisions at $\snn=2.76$ TeV using a set of triggers based on the hit multiplicity recorded by the VZERO detector, 
which consists of segmented scintillators covering the full azimuth over $2.8<\eta<5.1$ (VZERO-A) and $-3.7<\eta<-1.7$ (VZERO-C).
 
\section{Data analysis}
\label{sec:analysis}
A total of $11.5$M~($15\mu{\mathrm{b}}^{-1}$) and $5.7$M~($3.7\mu{\mathrm{b}}^{-1}$) events with VZERO multiplicities corresponding to $0$--$10$\% and $10$--$30$\%
most central events were selected using the centrality determination as described in \Ref{Abelev:2013qoq}.
The accepted events, reconstructed as described in \Ref{Abelev:2012hxa}, were required to have a primary reconstructed vertex within 
$\com{|{z}_{\rm vertex}|<}10$~cm of the center of the detector. 

Reconstructed tracks were required to have at least 3 hits in the ITS used in the fit to ensure adequate track momentum resolution for jet reconstruction. 
For tracks without any hit in the SPD, the primary vertex location was used in addition to the TPC and ITS hits for the momentum determination of the track.
This reduced the azimuthal dependence of the track reconstruction efficiency due to the non-uniform SPD response, 
without creating track collections with drastically differing momentum resolutions.  
Accepted tracks were required to be measured with $0.15<\pT<100$~\GeVc\ in $|\eta|<0.9$, 
and to have at least 70 TPC space-points and no less than 80\% of the geometrically findable space-points in the TPC.
The tracking efficiency was estimated from simulations of the detector response using GEANT3~\cite{geant3ref2} with the HIJING~\cite{hijing} event
generator as input. In $0$--$10$\% collisions, it is about 56\% at $0.15$~\GeVc, about $83$\% at $1.5$~\GeVc\ 
and then decreases to $81$\% at 3~\GeVc, after which it increases and levels off to about $83$\% at above 6.5~\GeVc.
In $10$--$30$\% collisions, the tracking efficiency follows a similar \pt{} dependence pattern, 
with absolute values of the efficiency that are $1$ to $2$\% higher compared to $0$--$10$\% collisions.
The momentum resolution $\delta\pT/\pT$, estimated on a track-by-track basis using the covariance matrix of the track fit, is about $1\%$ at $1.0$~\GeVc\ 
and about $3$\% at $50$~\GeVc. 
Tracks with $\pT>50$ \GeVc\ were only a small contribution to the inclusive jet population considered in this analysis,
for example only 20\% of the jets with \jetpt~larger than $100$~\GeVc~were found to contain a track above $50$~\GeVc.

EMCal cells with a calibrated response of more than 50 MeV were clustered prior to inclusion in the jet finder by a clustering algorithm 
which required each cluster to only have a single local maximum~\cite{Abelev:2014ffa}.
Interactions of slow neutrons or highly ionizing particles in the avalanche photodiodes create clusters with large apparent energy, 
but anomalously small number of contributing cells, and are removed from the analysis.
A non-linearity correction, derived from electron test beam data, of about $7$\% at $0.5$~GeV and negligible above $3$~GeV, was applied to the clusters' energies.
The energy resolution obtained from electron test beam data is about $15$\% at $0.5$~GeV and better than $5$\% above $3$~GeV.

Unlike electrons and photons, which deposit their full energy in the EMCal via electromagnetic showering, charged hadrons deposit energy 
in the EMCal, mostly via minimum ionization, but also via nuclear interactions which generate hadronic showers.
To avoid double counting, the energy deposited in the EMCal by charged particles that were already reconstructed as tracks,
the clusters' energies were corrected by the following procedure~\cite{Abelev:2013fn}:
All tracks with $\pT>0.15$~\GeVc\ were propagated to the average cluster depth within the EMCal, and then associated to 
clusters with $\ET>0.15$~GeV within the window $|\Delta\eta|<0.015$ and $|\Delta\varphi|<0.025$. 
Tracks were always matched to their nearest cluster, while clusters were allowed to have multiple track matches.  
Clusters with matched tracks were corrected for charged particle contamination by removing the fraction $f=100$\% of the sum of the momenta 
of all matched tracks from the cluster energy, as done in~\cite{Abelev:2013fn}.  
Clusters with $\ET>0.30$~GeV after this correction were used in this analysis.

The collection of tracks and corrected EMCal clusters was then assembled into jets using the anti-\kT~or the \kT~algorithms
in the \FastJet\ package~\cite{Cacciari:2011ma} with a resolution parameter of $R=0.2$.  
Only those jets that were at least $R$ away from the EMCal boundaries of $|\eta|<0.7$ and $1.4<\phi<\pi$, and thus fully contained 
within the EMCal acceptance, were kept in the analysis which limits the effect of the acceptance boundaries on the measured jet spectrum.
Jets reconstructed by the anti-\kT~algorithm were used to quantify signal jets,
while jets reconstructed by the \kT~algorithm were used to quantify the contribution from the underlying event.

\begin{figure}[tbh!f]
\begin{center}
 \includegraphics[width=0.75\textwidth]{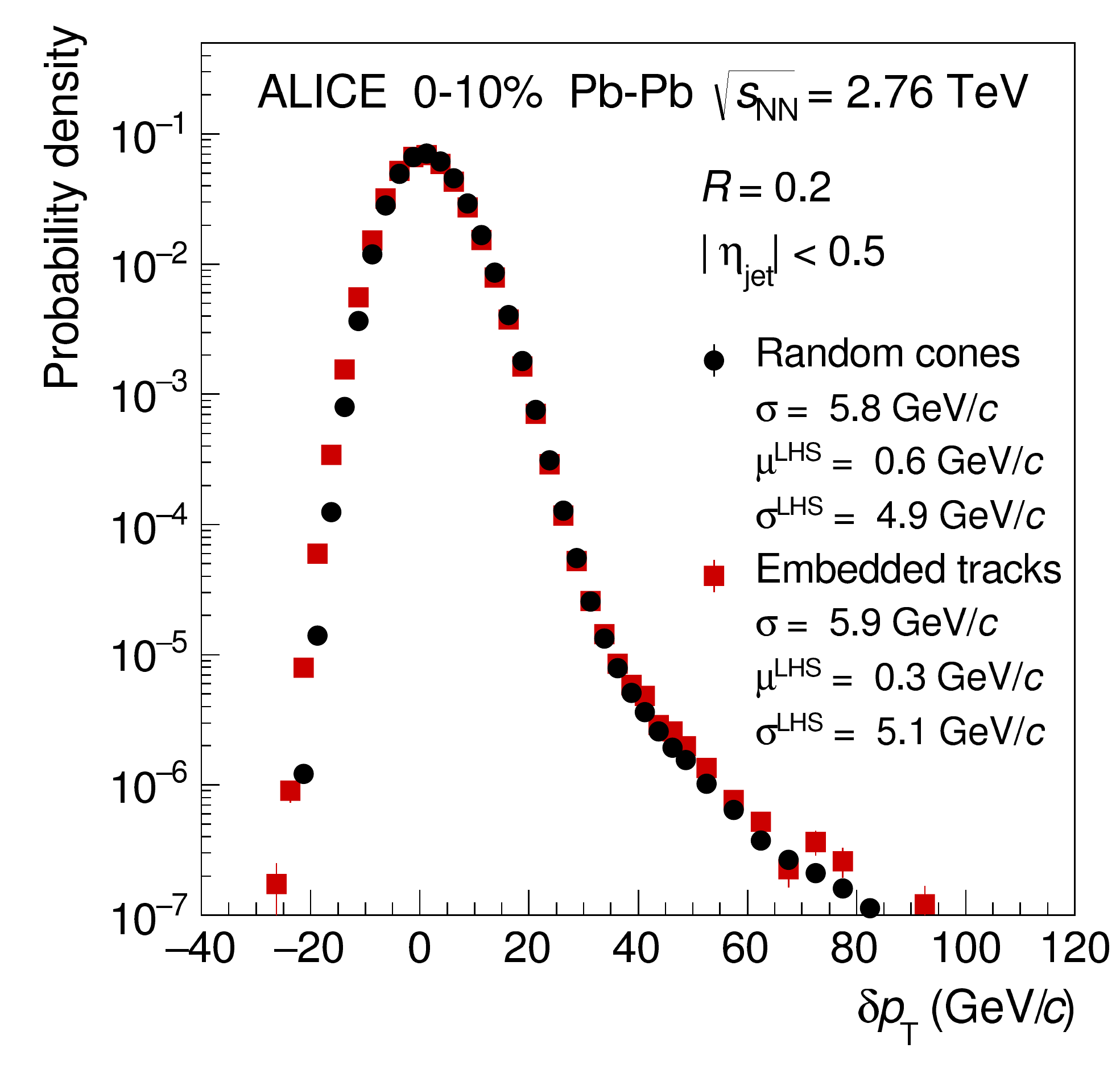}
 \caption{The $\dpT$ distribution for $R=0.2$ with the random-cone and the embedded-track methods in the 10\% most central events,
          with ${p}_{\rm T}^{\rm probe} = 60$~GeV/$c$ for the embedded-track method.} 
 \label{fig:DeltaPt}
\end{center}
\end{figure}

The signal spectrum formed from the reconstructed jets is affected by the contribution from the underlying event.  
In order to suppress the contribution of the background to the measurement of the jet energy, we followed the approach 
described in \Refs{Abelev:2013kqa,Abelev:2012ej}, which addresses the average additive contribution to the jet momentum on a jet-by-jet basis.
The underlying background momentum density was estimated event-by-event using the median of $\jetraw/\Ajet$, 
where \jetraw is the uncorrected energy and $\Ajet$ is the area of jets reconstructed with the \kT~algorithm.
Due to the limited acceptance of the EMCal, \rhoch, the median of the event-by-event momentum density distribution obtained 
from charged jets~(i.e.\ jets reconstructed from tracks only) in ${\abs{\jeteta}<0.5}$ and full azimuthal acceptance was used. \com{$30<\jetphi<230$}
Then, \rhochem\ was determined by scaling \rhoch\ using a centrality-dependent factor.
This factor is obtained from a parametrization of the measurement of the charged-to-neutral energy ratio,
using tracks and corrected clusters in the EMCal acceptance.
In $0$--$10$\% central \PbPb\ collisions, the average charged background momentum density was $\left<\rhoch\right>\approx110$~\GeVc. 
After scaling to include the neutral component we obtained $\left<\rhochem\right>\approx190$~\GeVc, which corresponds to an average 
contribution of the underlying event of about $24$~\GeVc\ in a cone of $R=0.2$.  
In $10$--$30$\% central \PbPb\ collisions $\left<\rhochem\right>$ decreases to $\approx130$~\GeVc{}.
For every signal jet reconstructed with the anti-\kT\ algorithm, the background density scaled by the area of the reconstructed 
signal jet was subtracted from the reconstructed transverse momentum of the signal jet according to $\jetreco=\jetraw-\rhochem\cdot\Ajet$.

Region-to-region background fluctuations lead to a smearing of the reconstructed jet energy.
Their magnitude was estimated as described in \Refs{Abelev:2013kqa,Abelev:2012ej} in two different ways: 
(1) by taking the scalar sum of the \pT\ of all particles found in a cone randomly placed in the event, referred to as random-cone method, and 
(2) embedding a single particle in the event and inspecting the anti-\kT\ jet that contains that embedded particle, referred to as embedded track method.
The first method does not rely on any assumptions about the structure of the background itself 
and gives approximately the same background fluctuation as embedding a track with infinite momentum for anti-\kT\ jets.
The second method should be able to reproduce the background as seen by the anti-\kT~algorithm more directly. 
The background fluctuations were quantified by $\dpT={p}_{\rm T}^{\rm cone} - \rhochem \cdot \pi{R}^{2}$ for the random-cone method,
and $\dpT= \jetreco - {p}_{\rm T}^{\rm probe}$ for the embedded-track method with a minimum of ${p}_{\rm T}^{\rm probe} = 10$~GeV/$c$ 
for the $\pt$ of the embedded track.
Above $10$~\GeVc\ the resulting $\dpT$ distribution does not depend on the $\pT$ of the embedded particle.
The \dpT\ distributions for the two methods in the 10\% most central collisions are shown in \Fig{fig:DeltaPt} for ${p}_{\rm T}^{\rm probe} = 60$~GeV/$c$.
The two methods appear to provide the same quantitative response to the background fluctuations, 
with only marginal differences mainly due to small jet area fluctuations in the embedding track method.
The widths of the $\dpT$~distributions are about 6 \GeVc.
The left-hand-side (LHS) of the distribution is Gaussian-like and is dominated by soft particle production.
To determine its width, the distributions were fitted recursively with a Gaussian function in the range 
$[\mu^{\rm LHS} - 3 \sigma^{\rm LHS}, \mu^{\rm LHS} + \frac{1}{2} \sigma^{\rm LHS}]$
using the mean and width of the $\dpT$~distribution as starting values for $\sigma$~and $\mu$.
The LHS width is about $5$~\GeVc\ in $0$--$10$\% and about $3.5$~\GeVc\ in $10$--$30$\% events.
The right-hand side has additional contributions from hard scattering processes, and the resulting non-Gaussian tail at high $\dpT$ is due to overlapping jets.
The random-cone method was used as the baseline in this analysis for creating the response matrix used in unfolding, while the single particle embedding method 
was used to study the sensitivity of the results to the method.

Additionally, signal jets were required to contain a charged track with a transverse momentum of at least $5$~\GeVc\ 
and a minimum background subtracted \jetreco~of $30$ \GeVc\ for $0$--$10$\% and of $20$ \GeVc\ for $10$--$30$\% most central events, 
which roughly corresponds to 5 $\sigma$ of the $\dpT$~distribution,
in order to suppress the contribution of combinatorial jets, i.e.\ from jets reconstructed mainly from upward fluctuations of the 
soft-particle background. 

Both the average background and the background fluctuations are averaged over all possible orientations of the event plane,
namely it is assumed that the signal jet sample being analyzed is isotropically distributed with respect to the event plane.
However, the jet sample may show some degree of correlation with the event plane, both for physical reasons (e.g.~path length dependence
of jet energy loss) or as a result of the cuts applied in the analysis (most notably the requirement on the leading hadron $\pT$). 
Since the background is also correlated with the event plane due to flow~($v_2$)~\cite{Aamodt:2010pa}, 
a question may arise about the validity of this approach.
Upper limits on the magnitude of these effects have been estimated by using random cones biased towards the event plane,
either by requiring the presence of a $5$~\GeVc\ track or by weighting the distribution using an upper limit on the jet $v_2$ of 0.1. 
In both cases, the upper limits on the shift of the jet energy scale~(JES) were found to be smaller than $0.1$~\GeVc.

\section{Unfolding}
\label{sec:unfolding}
The measured jet spectra are distorted by the response of the detectors used in the measurement 
and the background fluctuations in the underlying event.  
To correct for these effects we used an ``unfolding'' procedure, as described in \Ref{Abelev:2013kqa}.
The corrected distribution \jettrue\ and the measured distribution \jetreco\ are related by a convolution 
through the response matrix $\rm{RM_{tot}} = \rm{RM_{bkg}} \times \rm{RM_{det}}$, 
where $\rm{RM_{det}}$ parametrizes the detector response and $\rm{RM_{bkg}}$ the background fluctuations.
The unfolding procedure operates under the assumption that $\jetreco = \rm{{RM}_{tot}}\times \jettrue$.
Both background fluctuations and the detector response to jets are uniform within the $\eta$ and $\varphi$ acceptances, 
which is a precondition for the factorized approach used in building $\rm{RM_{tot}}$.

The detector response for jet reconstruction was obtained using \pp\ events simulated with the PYTHIA6~\cite{Sjostrand:2006za} event 
generator~(tune A~\cite{Skands:2010ak}).
Jets were reconstructed both at ``generator level'' and at ``detector level'' using the anti-\kT~algorithm.
Generator-level simulations utilized 
only prompt particles originating from the collision (with $c\tau < 1$~cm), directly 
from the event generator output, without accounting for detector effects;
detector-level simulations also included a detailed particle transport and detector response simulation based on GEANT3~\cite{geant3ref2}
with the detector response set to the \PbPb\ configuration.
During detector-level jet reconstruction, an additional $\pt$-dependent tracking inefficiency was introduced in order to account 
for the larger inefficiency due to the larger occupancy effects in central \PbPb\ events compared to \pp\ events.
Occupancy effects have been estimated comparing the tracking performance in PYTHIA and HIJING simulations, 
which represent \pp\ and \PbPb\ events \cite{hijing}.  
The occupancy effects in central HIJING events are larger for $\pT<0.5$~\GeVc\ 
where the efficiency is about $4$\% lower compared to PYTHIA, and then levels off to about $2$\% lower for $\pT>2$~\GeVc. 
In semi-central HIJING events, occupancy effects on the tracking efficiency amount to no more than $2$\% at low \pT\ and about $1$\% for $\pT>2$~\GeVc.
Other than this tracking efficiency correction, the detector response to jets was assumed to be the same in \PbPb\ events 
as in the PYTHIA simulated \pp~collisions.

The generator-level and detector-level jets were matched based on the Euclidean distance between their jet axes in pseudorapidity and azimuthal angle.  
It was ensured that the matching operation is bijective: each generator-level jet was matched to at most one detector-level jet~\cite{Abelev:2013kqa}.
Every matched jet pair corresponds to an entry in the detector response matrix, $\rm{RM_{det}}$.  
An unmatched generator-level jet represents a jet that was not reconstructed, and this distribution was used to determine the jet reconstruction efficiency.
In $0$--$10$\% \PbPb\ events, the detector jet reconstruction efficiency was found to be 90\% at $40$ \GeVc\ and 95\% above $70$~\GeVc, limited mainly by 
the track reconstruction efficiency of the leading charged particle.
As described above, at detector level the constituent cut was $150$~\MeVc\ for tracks, and $300$~MeV for clusters after the cluster energy
is corrected for charged particle energy contamination. 
However, at generator level no such cut is applied, and hence the reconstructed jets are corrected to a constituent charged particle momentum of
$0$~\MeVc\ and to a constituent cluster energy of $0$~MeV in the unfolding process.  
A net negative shift of the JES at detector level was obtained, which originates mainly from tracking inefficiency 
and unreconstructed particles, such as neutrons and $K^0_{\rm L}$, though the subtraction procedure for energy deposits 
by charged particles in the EMCal and missing secondary particles from weak decays contribute to the shift \cite{Abelev:2013fn}.
The JES correction applied through the response matrix is about 23\% at $\jettrue$ of $40$ \GeVc\ and 29\% at $120$ \GeVc\,
independent of centrality.

The $\rm{RM_{bkg}}$ matrix was constructed row-by-row by taking the $\dpT$ distribution and shifting it along the $\jetreco$ axis 
by the amount $\jettrue$ corresponding to each row (Toeplitz matrix).
This matrix construction method assumes that the response of the jet spectrum to background fluctuations is independent of the jet momentum.

The \pT-dependence of the jet momentum resolution $\sigma(\jetreco) / \jettrue$ is different for the background and detector contributions \cite{Abelev:2013kqa}.
The contribution from background fluctuations is dominant at low $\jettrue$ and is proportional to $1/\jettrue$, 
whereas the contribution from detector effects is fairly constant with $\jettrue$. 
The cross-over between the two contributions happens at $\jettrue\approx30$~\GeVc. 
The combined jet momentum resolution is about 23\% at $\jettrue$ of $40$ \GeVc\ and 20\% at $120$~\GeVc\ for $0$--$10$\% collisions, 
while it is 24\% at $\jettrue$ of $30$ \GeVc\ and 20\% at $100$~\GeVc\ for $10$--$30$\%.

Two unfolding algorithms with different regularization procedures were used for correcting the measured jet spectrum: 
the ${\chi}^{2}$ minimization method~\cite{Verkerk:1985hw} with a log-log-regularization and the generalized Singular Value Decomposition (SVD) 
method~\cite{Hocker:1995kb}, as implemented in RooUnfold~\cite{Prosper:1306523}, 
which was used for the default value of the data points.  
The measured spectrum used as an input to the unfolding was in the range $30<\jetpt<120$~\GeVc\ for $0$--$10$\% and $20<\jetpt<100$~\GeVc\ 
for $10$--$30$\% collisions.
A smoothed version of the measured spectrum was used as the prior, so that the statistical fluctuations within the data were not 
magnified in the unfolding process. 
The regularization parameter used for SVD unfolding is $k=5$. 
The value of $k$ is chosen such that it corresponds to the $d$ vector magnitude of 1, and Pearson coefficients 
which do not show a large variation in the correlation between neighboring \pt~bins. 

The corrected jet spectra are reported for $40<\jetpt<120$~\GeVc\ in $0$--$10$\%, and for $30<\jetpt<100$~\GeVc\ in $10$--$30$\% 
where the efficiency due to these kinematic cuts\com{, called the kinematic efficiency,} is high, approximately 90\%.
It was verified that the cut on the reconstructed jet \pT\ has a negligible effect 
in the reported \pT\ region of the final result, as long as the requirement on the leading charged track \pT\ is at least $5$~\GeVc{}.
If this threshold is reduced, the cut on the minimum reconstructed jet \pT\ becomes crucial for unfolding stability.

The analysis procedures in the $10$\% most central collisions were tested with two different Monte Carlo~(MC) models, where events were 
constructed by embedding jets into a soft background. 
The first test verified the robustness of the unfolding framework with the inclusion of fake ``jets'' that are clustered from the soft background, 
which did not originate from a hard process.  
The second model tested the assumption that the background and detector responses can be factorized. 

In the first model, the soft background of both charged and neutral particles was modeled with $3100 < {N}_{\rm tracks} < 5150$ 
where the particle transverse momenta were taken from a Boltzmann distribution with a temperature of 550 MeV. 
This model created a fluctuating background similar to that of the 0-10\% \PbPb\ data; 
e.g.\ the background fluctuations, as estimated via the
\dpT\ distributions, coincide within few percent.
Jets were reconstructed at generator level in PYTHIA-only events and at detector level, with the added background. 
The first model validated the background subtraction technique, and in particular the stability of the unfolding method
against the contribution from the residual combinatorial background.
In the second model, the background was taken from real $0$--$10$\% \PbPb\ events. 
The charged particle correction for the EMCal clusters was applied after embedding.
Only jets with at least 1 \GeVc\ of transverse momentum coming from the embedded PYTHIA event were selected for the test. 
This is needed to reject the signal from hard scatterings in the data, but also removes most of the combinatorial jets from the \PbPb\ underlying event.
The second model was used to test the validity of the charged particle correction applied to the EMCal clusters, 
in particular in the interplay between the underlying event and the jets.
It also validates certain aspects of the corrections applied for the background fluctuations, e.g.\ the
unsmearing of the jet \pT\ due to background fluctuations or the overlap with low momentum jets.
Background tracks and clusters could be matched to jet tracks and clusters or vice versa, so that
the correction for charged particle contamination could potentially cause an over-subtraction 
that is not corrected for in the unfolding procedure.
These Monte Carlo tests showed that the analysis procedures outlined above, including unfolding, 
recovered the input spectrum within the statistical and systematic uncertainties of the models.

\begin{table}[htb!f]
\begin{center}
\begin{tabular}{lrrrrr} Category & \multicolumn{5}{c}{Relative uncertainty (\%)}  \\
   \hline
   \hline
                              &  $\jetpt^{\rm min}$ &  $\jetpt^{\rm max}$ &  Min.     &  Max.     &  Avg.      \\
Tracking efficiency           &  7.7     &  11.3    &  7.3     &  11.3    &  8.8       \\
Track momentum resolution     &  1.0     &  1.0     &  1.0     &  1.0     &  1.0       \\
Charged particle correction   &  0.7     &  2.7     &  0.7     &  6.4     &  3.7       \\
EMCal clusterizer             &  3.2     &  1.8     &  0.1     &  3.2     &  1.4       \\
EMCal response                &  4.4     &  4.4     &  4.4     &  4.4     &  4.4       \\
Background fluctuations       &  3.9     &  2.7     &  2.3     &  3.9     &  2.8       \\
Jet raw $\pt$ cuts            &  2.6     &  6.7     &  1.5     &  6.7     &  3.6       \\
Combinatorial jets            &  0.3     &  0.5     &  0.0     &  0.5     &  0.2       \\
\hline
Total correlated uncertainty  &  10.6    &  14.5    &  10.6    &  14.5    &  12.0      \\
\hline
\hline
Unfolding method              &  0.1     &  10.0    &  0.1     &  15.5    &  6.6       \\
SVD reg. param. $k = 4$       &  3.6     &  11.7    &  2.4     &  11.7    &  6.0       \\
SVD reg. param. $k = 6$       &  7.2     &  2.7     &  1.5     &  8.8     &  5.3       \\
Prior choice 1                &  1.9     &  4.0     &  0.2     &  4.0     &  1.6       \\
Prior choice 2                &  2.1     &  1.4     &  0.1     &  2.1     &  0.9       \\
\hline
Total shape uncertainty       &  3.8     &  7.2     &  2.7     &  7.4     &  5.3       \\
\hline
\hline
\end{tabular}
\caption{Summary of systematic uncertainties for $0$--$10$\% most central collisions. 
         The first column is the uncertainty at the minimum $\jetpt^{\rm min}$ of 40~\GeVc, 
         the second column is the uncertainty at the maximum $\jetpt^{\rm max}$ of 120~\GeVc.  
         The minimum and maximum columns give the extreme, and the last column gives 
         the average systematic uncertainty over the entire $\pt$ range.
         The total correlated uncertainty was calculated by adding the components in quadrature, 
         while the shape uncertainty was calculated as the $\sigma$ of the different variations 
         (see text for details).
}
\label{tab:syst}
\end{center}
\end{table}

\begin{figure}[tbh!f]
\begin{center}
 \includegraphics[width=0.75\textwidth]{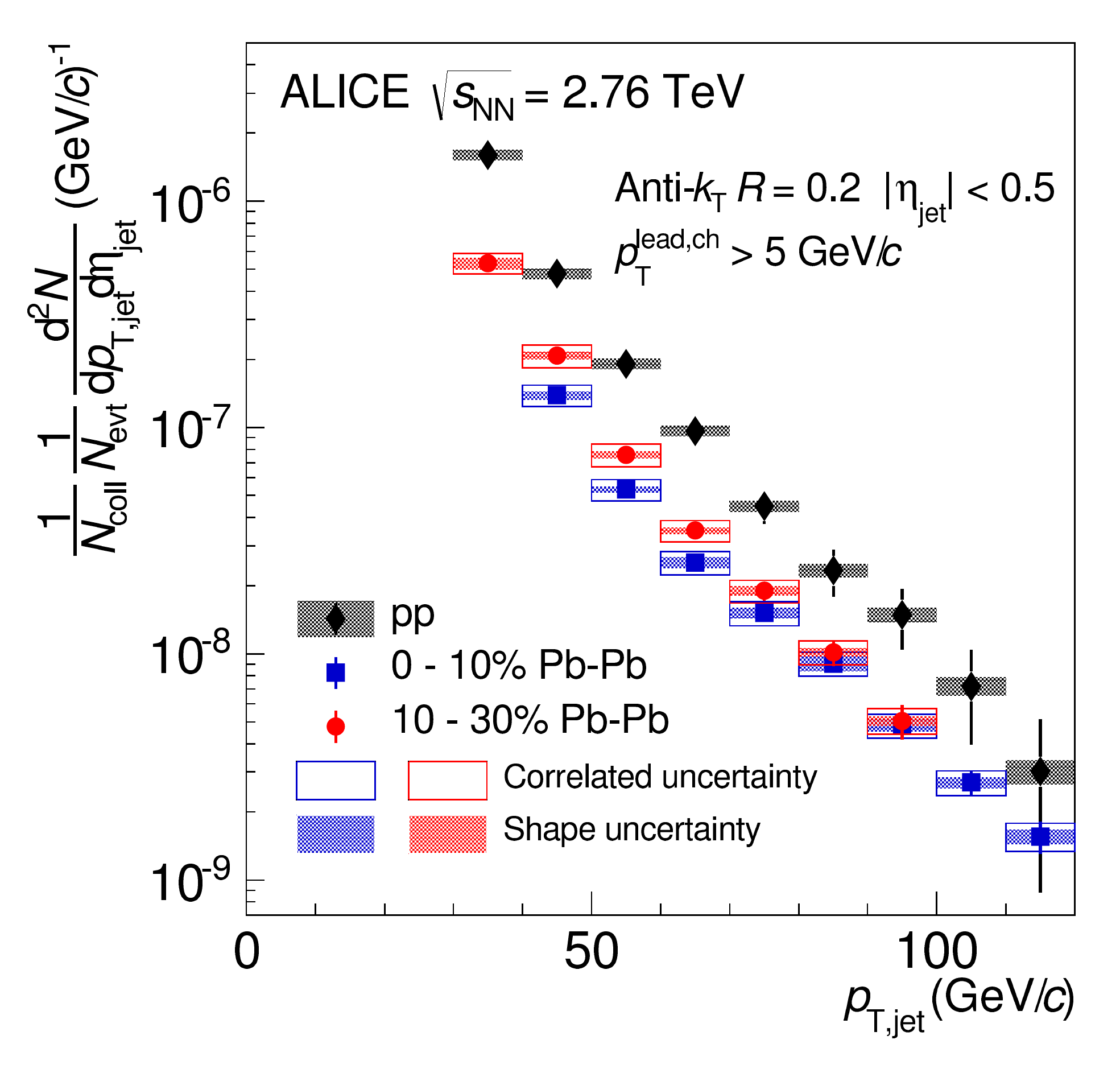}
 \caption{The spectra of $R=0.2$ jets with a leading track requirement of $5$~\GeVc\ in $0$--$10$\% and $10$--$30$\% most 
          central \PbPb\ collisions scaled by $1/\Ncoll$ and in inelastic \pp\ collisions at $\snn=2.76$~TeV. The uncertainties on the 
          normalization are about 11\% for the \PbPb\ data from the uncertainty on \Ncoll\ and about 8\% for the \pp\ data from the 
          total inelastic cross section.}
 \label{fig:PbPbSpectraR02Bias5}
\end{center}
\end{figure}

\section{Results}
\label{sec:results}
The unfolded jet spectra in 0$-$10\% and 10$-$30\% central collisions are displayed in \Fig{fig:PbPbSpectraR02Bias5}.
To compare the spectra with the spectrum measured in \pp\ collisions, the yield is divided by the number of binary collisions,
which is $\Ncoll=1501\pm167$\ for $0$--$10$\% and $743\pm79$ for $10$--$30$\% collisions, 
as estimated from a Glauber MC calculation~\cite{Abelev:2013qoq}. 

The systematic uncertainties on the jet spectrum are summarized in Tab.~\ref{tab:syst} for the 0-10\% centrality class.
For the 10-30\% centrality class the corresponding uncertainties differ, on average, by 2\% or less.
The systematic uncertainties were divided into two categories: correlated uncertainties and shape uncertainties.
The correlated uncertainties result dominantly from uncertainties on the JES, 
such as the uncertainty of the tracking efficiency, that will shift the entire jet 
spectrum in one direction, whereas the shape uncertainties are related to the unfolding and can distort the slope of the spectrum.  

The dominant correlated uncertainty on the jet spectrum of about 9\% arises from the uncertainty on the tracking efficiency.
It is estimated by varying the tracking efficiency by 5\% in determining $\rm{RM_{det}}$ and unfolding the spectrum.
The uncertainty due to the correction procedure for the charged particle double counting in the EMCal of about 
$4$\% was determined by varying $f$ from 100\% \com{to 70\% and} to 30\% in both the measured spectrum and the $\rm{RM_{det}}$. 
The determination of the uncertainties from other EMCal response related uncertainties as EMCal energy scale, EMCal energy resolution, 
and EMCal non-linearity is outlined in~\cite{Abelev:2013fn} and combined leads to an uncertainty of 4.4\%.
The uncertainty arising from the choice of the EMCal clustering algorithm is determined by using a different clusterizing method, 
that forms fixed-size clusters from 3x3 towers. 
For the background fluctuations, the response matrix $\rm{RM_{bkg}}$~was constructed with the single-track embedding method
for determining \dpT, as discussed above.
To estimate the sensitivity of the unfolding to the raw jet $\pt$ selection, the \pT\ range of input spectra is varied by extending
the range at both the low and high ends by $\pm5$~\GeVc.
The influence of combinatorial jets, estimated by varying the low edge of the unfolded spectrum from $0$ to up to $10$~\GeVc\, was found to be negligible.
Since all sources of uncertainty are independent, each contribution is added in quadrature to obtain the final correlated uncertainty 
of 10.6\% to 14.5\% as listed in Tab.~\ref{tab:syst}. 
The uncertainty on the JES is 2.4\% to 3.2\% and can be obtained by dividing the uncertainties listed in Tab.~\ref{tab:syst} by $4.5$, 
where the exponent $n=4.5$ was obtained by fitting a power law to the measured spectrum.

\begin{figure}[tbh!f]
\begin{center}
 \includegraphics[width=0.75\textwidth]{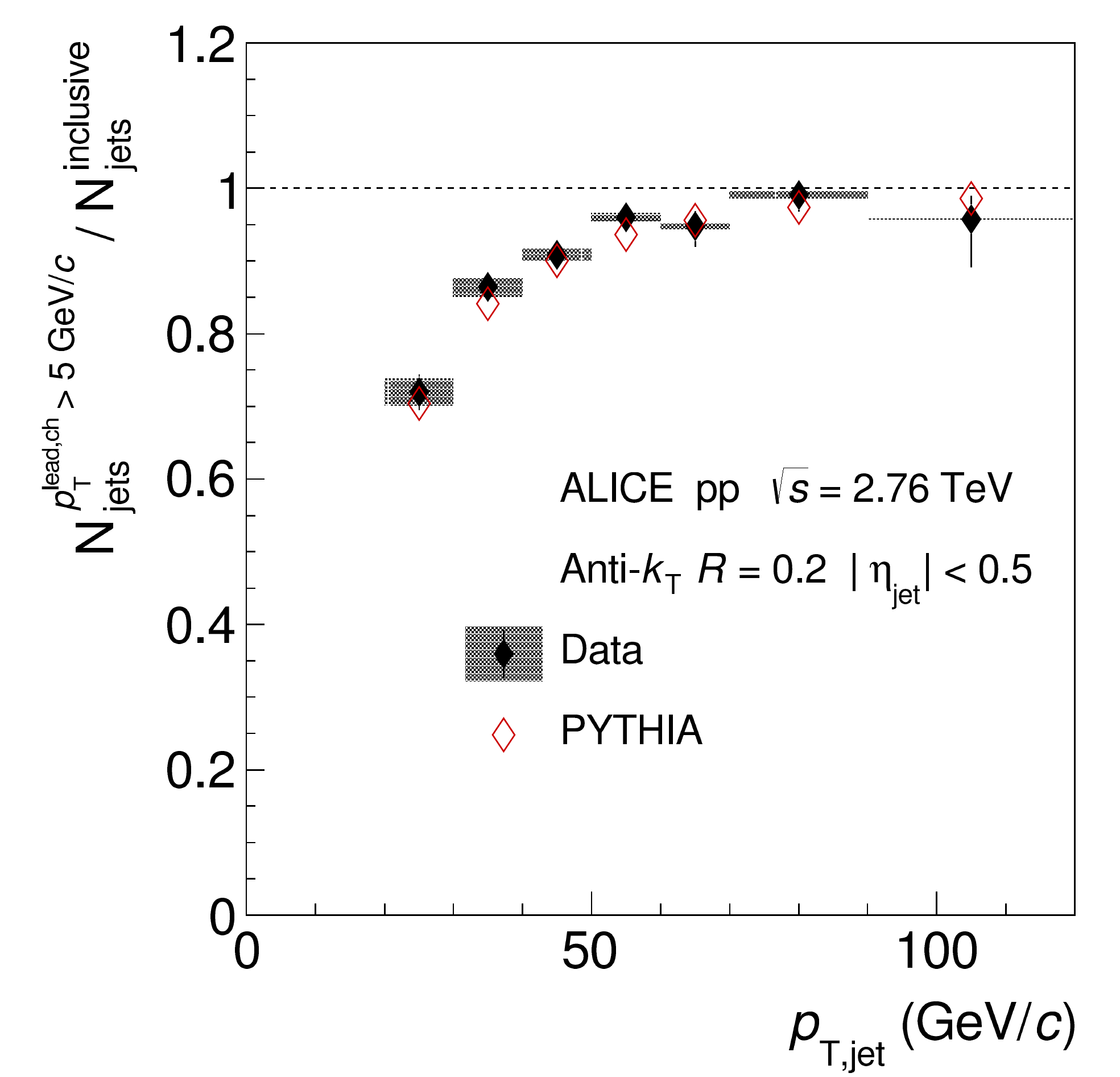}
 \caption{Ratio of the jet spectrum with a leading track $\pT > 5$~\GeVc\ over the inclusive jet spectrum for $R=0.2$ in \pp\ collisions at $\s=2.76$~TeV.}
 \label{fig:ppRatioR02}
\end{center}
\end{figure}

The shape uncertainty is dominated by the regularization used in the unfolding and can be divided into two components: 
the method by which the solution is regularized, e.g.\ ${\chi}^{2}$ instead of the SVD unfolding,
and the variation of the regularization process within a given method. 
The regularization is done by adding a penalty term in the ${\chi}^{2}$ method and by ignoring the components of the SVD decomposition 
that are dominated by statistical fluctuations.
For the SVD method, the regularization $k$ factor is an integer value and thus can only be varied in integer steps.  
The uncertainty related to the choice of the prior is estimated by varying the exponent of the power law function 
extracted from the reconstructed spectrum by $\pm0.5$, which is used to construct the prior. 
The uncertainty related to the choice of the prior is estimated by varying the exponent $n=4.5$ by $\pm0.5$ to scale the prior.
The differences in the unfolded spectrum with these variations are summarized in Tab.~\ref{tab:syst}.  
These variations in the regularization strategy are combined assuming that they constitute independent measurements. 
The final shape uncertainty is thus obtained by summing them in quadrature and dividing by the square root of the number of variations.
 
The jet spectrum in \pp\ collisions\com{ at $\s=2.76$~TeV} was measured in the same way as reported in \Ref{Abelev:2013fn}, but with the $5$~\GeVc\ 
leading charged particle requirement necessary for the \PbPb\ analysis. The resulting spectrum normalized per inelastic \pp\ collision 
is shown in \Fig{fig:PbPbSpectraR02Bias5}.
In order to determine the effect of the leading track requirement in \pp\ collisions, the ratio of the jet spectra with a $5$~\GeVc\ 
leading track requirement (the biased jet sample), over the spectrum of jets without a leading track requirement (the inclusive jet sample) 
with resolution parameter $R=0.2$ is shown in \Fig{fig:ppRatioR02}.
Systematic uncertainties in the ratio were evaluated by removing the uncertainties 
that are correlated between the spectra obtained with and without the cut on the leading particle.
As can be seen in~\Fig{fig:ppRatioR02}, for \jetpt\ above $50$~\GeVc\, more than $95$\% of all reconstructed jets have at least one track 
with a \pT~greater than 5 \GeVc. 
PYTHIA tune A (but also other common tunes like the Perugia tunes~\cite{Skands:2010ak}) accurately describe the measured ratio. 

\begin{figure}[tbh!f]
\begin{center}
 \includegraphics[width=0.99\textwidth]{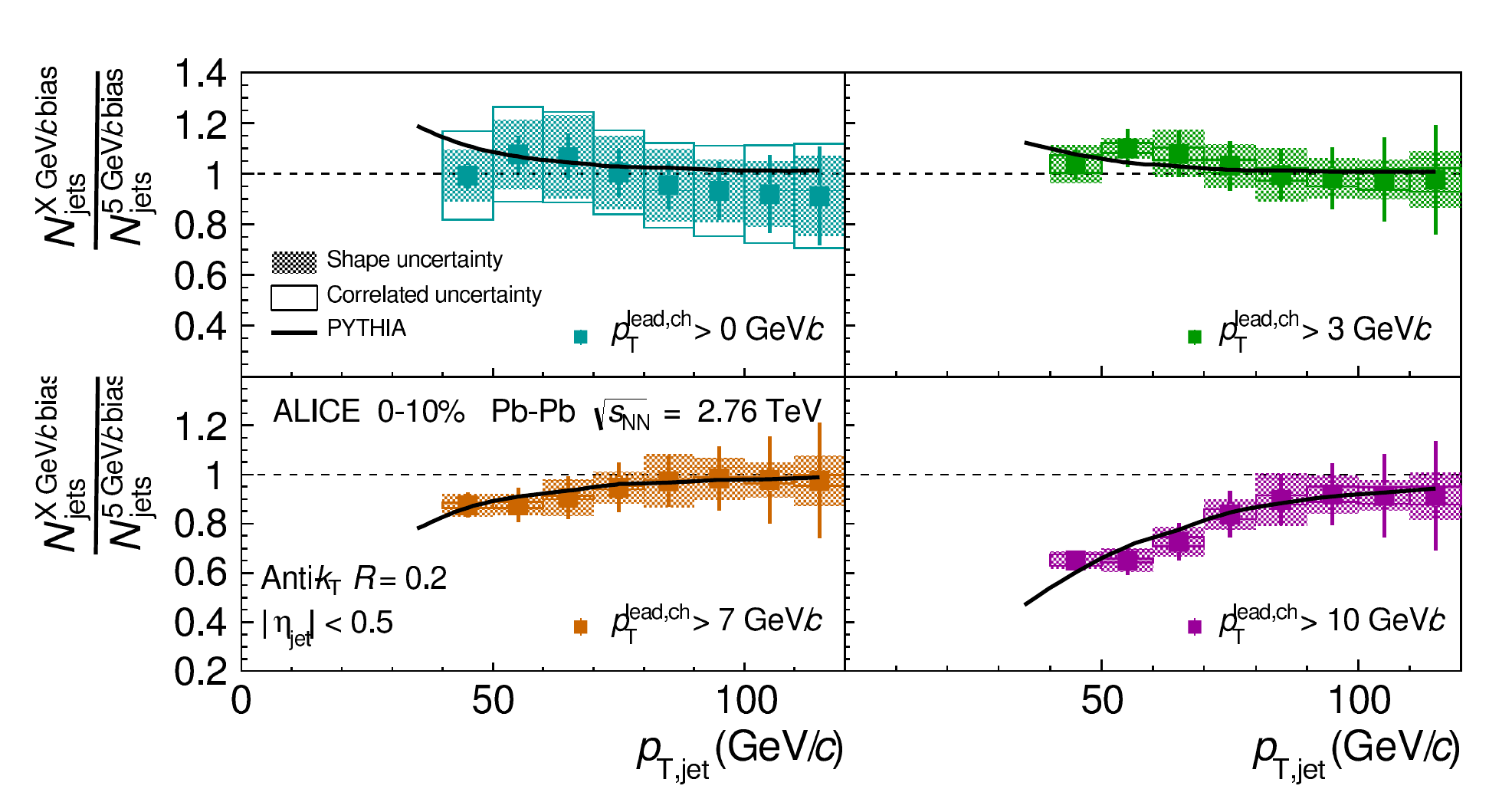}
 \caption{Ratios of jet spectra with different leading track $\pT$ requirements~(``$0$ over $5$'', ``$3$ over $5$'', ``$7$ over $5$''and ``$10$ over $5$'') 
          for $R=0.2$ jets in 0-10\% \PbPb\ collisions at $\snn=2.76$~TeV.  The solid black lines represent predictions from PYTHIA.}
 \label{fig:BiasRatio}
\end{center}
\end{figure}

The influence of the leading track requirement in the \PbPb\ measurement, nominally set to $5$~\GeVc\, was tested by varying it by $40$\%, 
i.e.\ reducing it to $3$ and increasing it to $7$~\GeVc, and with the more extreme values of $0$ and $10$~\GeVc. 
The ratios of jet spectra with the different leading track $\pT$ biases, after all corrections, are shown in \Fig{fig:BiasRatio} 
for $R=0.2$ jets in 0-10\% central \PbPb\ collisions at $\snn=2.76$~TeV.  The corrections to these different jet spectra were done using 
the same unfolding procedure as the nominal spectrum with leading track $\pT$ bias of 5~\GeVc, with a slightly modified response matrix which accounts 
for the different biases.
Since the unfolding procedure weakens the correlation between the statistical fluctuations of the jet spectra with different
leading track requirements, the statistical uncertainties have been added in quadrature in the ratio. 
The systematic shape uncertainty is due to the unfolding procedure, and has been treated as completely uncorrelated in the ratio. 
The correlated uncertainty is primarily due to the uncertainty on the JES, which is highly correlated between the various spectra. 
The systematic variations in the unfolding procedure have been applied consistently for both the denominator (with a leading track $\pT > 5$~\GeVc) 
and the numerators (with a $0$, $3$, $7$ and $10$~\GeVc\ leading track bias), and the resulting difference in the ratios has been 
taken as a systematic uncertainty.
The jet spectra with leading track requirements of $3$ and $0$~\GeVc\ are consistent with the baseline measurement with a $5$~\GeVc\ requirement.  
The unfolding is not as stable as with a $5$~\GeVc\ requirement, which leads to a 
larger systematic uncertainty due to the unfolding correction procedure, 
especially for the inclusive spectrum.
All measurements of the ratio of jet spectra with different leading track biases, particularly those with a higher leading track $\pT$ 
requirement than the nominal, are well described by PYTHIA 6 (tune A), within one sigma of the uncertainties or less.

\begin{figure}[tbh!f]
\begin{center}
 \includegraphics[width=0.9\textwidth]{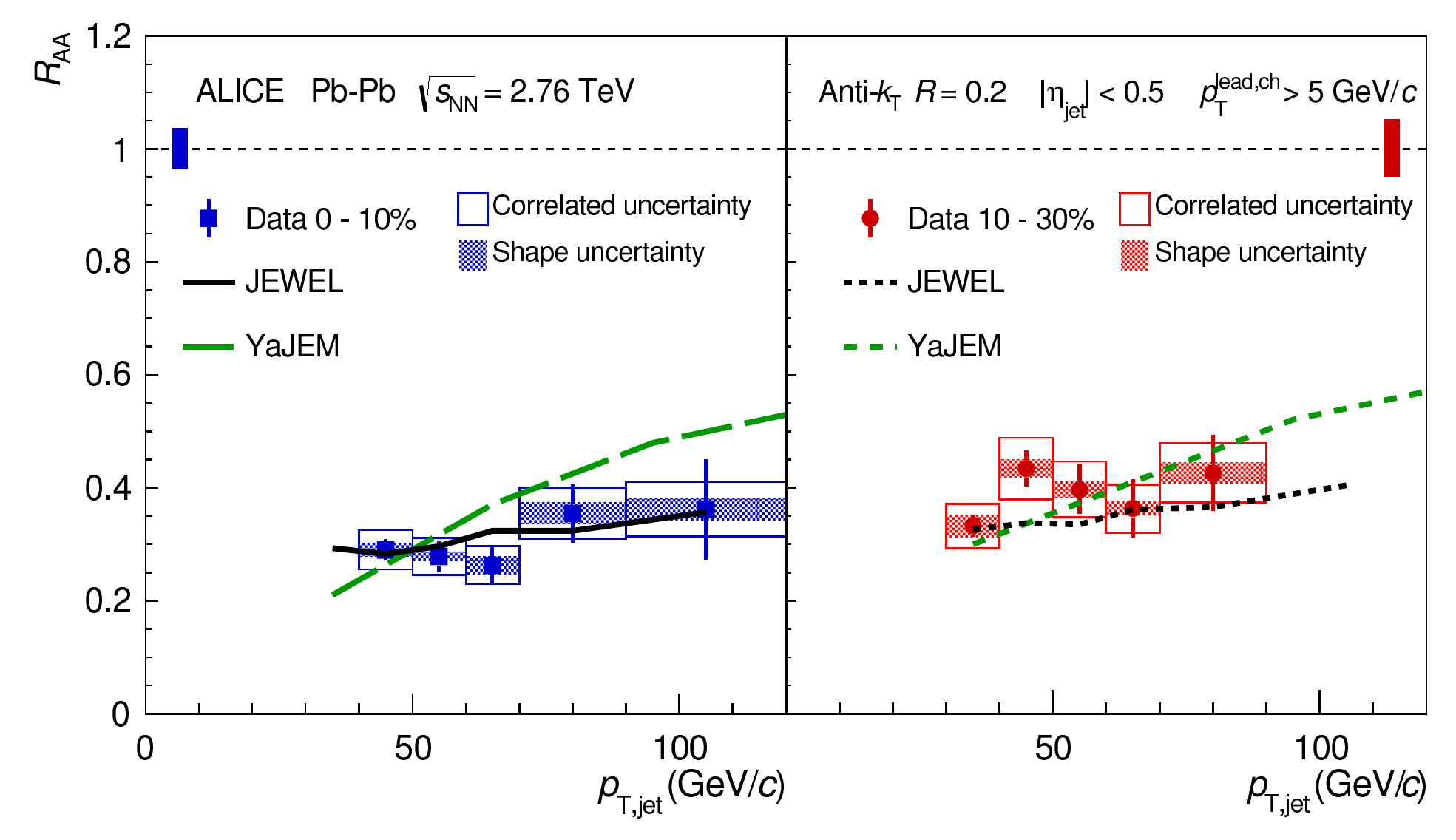}
 \caption{$\RAA$ for $R=0.2$ jets with the leading track requirement of $5$~\GeVc\ in $0$--$10$\% (left) and $10$--$30$\% (right) 
          most central \PbPb\ collisions compared to calculations from YaJEM~\cite{Renk:2013rla} 
          and JEWEL~\cite{Zapp:2012ak}.
          The boxes at $\RAA=1$ represent the systematic uncertainty on $\TAA$.} 
 \label{fig:RAATheoryCompare}
\end{center}
\end{figure}

The nuclear modification factor, \RAA, is defined as the ratio of the jet spectrum in \PbPb\ divided by the spectrum in \pp\ collisions scaled by \Ncoll.
It is constructed such that \RAA\ equals unity if there is no net nuclear modification of the spectrum in \PbPb\ collisions 
as compared to an incoherent superposition of independent \pp~collisions. 
The resulting $\RAA$ of jets with a $5$~\GeVc\ leading track requirement for $R=0.2$ in the 0$-$10\% and 10$-$30\% central \PbPb\ 
collisions is reported in \Fig{fig:RAATheoryCompare}.
The systematic and statistical uncertainties from the \PbPb\ and \pp\ measurements (see \Fig{fig:PbPbSpectraR02Bias5}) are added in quadrature.
The resulting uncertainty on the normalization is from scaling the \pp\ cross section with the nuclear overlap $T_{\rm AA}=23.5 \pm 0.87$~mb$^{-1}$ for $0$--$10$\% 
and $11.6 \pm 0.60$~mb$^{-1}$ for $10$--$30$\% collisions. 
As can be seen, jets in the measured $\jetpt$ range are strongly suppressed. 
The average \RAA\ in both $0$--$10$\%~and $10$--$30$\% central events was found to have a negligible $\jetpt$~dependence.
In the 10\% most central events, combining the statistical and systematic uncertainty in quadrature, the average \RAA\ is found to be $0.28\pm0.04$.
The suppression is smaller in magnitude in the $10$--$30$\% central events, leading to an average \RAA of $0.35\pm0.04$. 
These results qualitatively agree with the suppression obtained from measurements using charged-particle jets~\cite{Abelev:2013kqa}, 
though the jet energy scale is not the same in both cases, and so a direct comparison is not possible.  
Furthermore, the results are consistent with the \RAA~reported by ATLAS for $R$ = 0.4 jets scaled by the ratio of the yields 
with the different resolution parameters in different \jetpt~bins~\cite{Aad:2014bxa,Aad:2012vca}. 

In order to interpret the results and move to a more quantitative understanding of jet quenching mechanisms, a comparison of the measured \RAA\
in $0$--$10$\% central collisions to calculations from two different models is also shown in~\Fig{fig:RAATheoryCompare}. 
The first model, YaJEM~\cite{Renk:2013rla}, uses a 2+1D hydrodynamical calculation and a Glauber MC for the initial geometry\com{ of nucleon--nucleon 
collisions}, as well as a LO pQCD calculation to determine the outgoing partons.  
Parton showers are modified by a medium-induced increase of the virtuality during their evolution through the medium.
The Lund model in PYTHIA is used for hadronization into final state particles. 
The kinematics of the virtual partons in the evolving partonic shower were modified with a parameter related to the two transport coefficients, 
\qhat~and $\hat{e}$, that describes how strongly a parton of a given momentum couples to the medium. 
The parameter was fixed so that the model accurately describes the \RAA~for charged hadrons at 10 \GeVc~\cite{Chatrchyan:2011sx}, 
but no additional changes were made for the prediction of the jet \RAA. 
The second model, JEWEL~\cite{Zapp:2012ak}, takes a different approach in the description of the parton--medium interaction 
by giving a microscopical description of the transport coefficient, \qhat.  
Essentially each scattering of the initial parton with medium partons is computed and the average over all scatters determines \qhat.  
JEWEL uses a combination of Glauber and PYTHIA to determine the initial geometry, 
a 1D Bjorken expansion for the medium evolution, and PYTHIA for hadronization into final state particles.
The transverse medium density profile in JEWEL is proportional to the density of wounded nucleons combined with a 1D Bjorken expansion for the time evolution. 
Hard scatters are generated according to Glauber collision geometry, and suffer from elastic and radiative energy loss in the medium, including a Monte Carlo 
implementation of LPM interference effects. 
PYTHIA is used for the hadronisation of final state particles.
Despite their different approaches, both calculations are found to reproduce the jet suppression.
YaJEM, however, exhibits a slightly steeper increase with jet $\pt$ than the data.
The calculated $\chi^2$ are $1.690$ for YaJEM and $0.368$ for JEWEL, obtained by comparing the models with the data.
Additional measurements will be needed in order to further constrain the models, such as measuring the jet suppression 
relative to the event plane angle, which would require a more accurate modeling of the path-length dependence of jet quenching.

\section{Summary}
\label{sec:summary}
The transverse momentum~(\pT) spectrum and nuclear modification factor~(\RAA) of jets reconstructed from charged particles 
measured by the ALICE tracking system and neutral energy measured by the ALICE Electromagnetic Calorimeter are measured
with $R=0.2$ in the range of $40<\jetpt<120$~\GeVc\ for $0$--$10$\% and in $30<\jetpt<100$ \GeVc\ for $10$--$30$\% 
most central \PbPb\ collisions at $\snn=2.76$~TeV were measured. 
The jets were required to contain at least one charged particle with $\pT>5$~\GeVc. 
The effect of this requirement on the reported $\RAA$ was evaluated by the ratios of the jet spectra with the 5 GeV/$c$
to no requirement compared to expectations on PYTHIA, and found not to have an observable effect within the uncertainties
of the measurement.
Jets with $40<\jetpt<120$~\GeVc\ are strongly suppressed in the 10\% most central events, with \RAA\ about $0.28\pm0.04$, 
independent of $\jetpt$ within the uncertainties of the measurement.
The suppression in $10$--$30$\% events is $0.35\pm0.04$, slightly less than in the most central events.
The observed suppression is in fair agreement with expectations from two jet quenching model calculations.

%% file: acknowledgements_jan2015.tex
The ALICE Collaboration would like to thank all its engineers and technicians for their invaluable contributions to the construction of the experiment and the CERN accelerator teams for the outstanding performance of the LHC complex.
The ALICE Collaboration gratefully acknowledges the resources and support provided by all Grid centres and the Worldwide LHC Computing Grid (WLCG) collaboration.
The ALICE Collaboration acknowledges the following funding agencies for their support in building and
running the ALICE detector:
State Committee of Science,  World Federation of Scientists (WFS)
and Swiss Fonds Kidagan, Armenia,
Conselho Nacional de Desenvolvimento Cient\'{\i}fico e Tecnol\'{o}gico (CNPq), Financiadora de Estudos e Projetos (FINEP),
Funda\c{c}\~{a}o de Amparo \`{a} Pesquisa do Estado de S\~{a}o Paulo (FAPESP);
National Natural Science Foundation of China (NSFC), the Chinese Ministry of Education (CMOE)
and the Ministry of Science and Technology of China (MSTC);
Ministry of Education and Youth of the Czech Republic;
Danish Natural Science Research Council, the Carlsberg Foundation and the Danish National Research Foundation;
The European Research Council under the European Community's Seventh Framework Programme;
Helsinki Institute of Physics and the Academy of Finland;
French CNRS-IN2P3, the `Region Pays de Loire', `Region Alsace', `Region Auvergne' and CEA, France;
German Bundesministerium fur Bildung, Wissenschaft, Forschung und Technologie (BMBF) and the Helmholtz Association;
General Secretariat for Research and Technology, Ministry of
Development, Greece;
Hungarian Orszagos Tudomanyos Kutatasi Alappgrammok (OTKA) and National Office for Research and Technology (NKTH);
Department of Atomic Energy and Department of Science and Technology of the Government of India;
Istituto Nazionale di Fisica Nucleare (INFN) and Centro Fermi -
Museo Storico della Fisica e Centro Studi e Ricerche "Enrico
Fermi", Italy;
MEXT Grant-in-Aid for Specially Promoted Research, Ja\-pan;
Joint Institute for Nuclear Research, Dubna;
National Research Foundation of Korea (NRF);
Consejo Nacional de Cienca y Tecnologia (CONACYT), Direccion General de Asuntos del Personal Academico(DGAPA), M\'{e}xico, :Amerique Latine Formation academique – European Commission(ALFA-EC) and the EPLANET Program
(European Particle Physics Latin American Network)
Stichting voor Fundamenteel Onderzoek der Materie (FOM) and the Nederlandse Organisatie voor Wetenschappelijk Onderzoek (NWO), Netherlands;
Research Council of Norway (NFR);
National Science Centre, Poland;
Ministry of National Education/Institute for Atomic Physics and Consiliul Naţional al Cercetării Ştiinţifice - Executive Agency for Higher Education Research Development and Innovation Funding (CNCS-UEFISCDI) - Romania;
Ministry of Education and Science of Russian Federation, Russian
Academy of Sciences, Russian Federal Agency of Atomic Energy,
Russian Federal Agency for Science and Innovations and The Russian
Foundation for Basic Research;
Ministry of Education of Slovakia;
Department of Science and Technology, South Africa;
Centro de Investigaciones Energeticas, Medioambientales y Tecnologicas (CIEMAT), E-Infrastructure shared between Europe and Latin America (EELA), Ministerio de Econom\'{i}a y Competitividad (MINECO) of Spain, Xunta de Galicia (Conseller\'{\i}a de Educaci\'{o}n),
Centro de Aplicaciones Tecnológicas y Desarrollo Nuclear (CEA\-DEN), Cubaenerg\'{\i}a, Cuba, and IAEA (International Atomic Energy Agency);
Swedish Research Council (VR) and Knut $\&$ Alice Wallenberg
Foundation (KAW);
Ukraine Ministry of Education and Science;
United Kingdom Science and Technology Facilities Council (STFC);
The United States Department of Energy, the United States National
Science Foundation, the State of Texas, and the State of Ohio;
Ministry of Science, Education and Sports of Croatia and  Unity through Knowledge Fund, Croatia.
Council of Scientific and Industrial Research (CSIR), New Delhi, India

%% file: Alice_Authorlist_2015-Jan-29.tex


\begingroup
\small
\begin{flushleft}
J.~Adam\Irefn{org39}\And
D.~Adamov\'{a}\Irefn{org82}\And
M.M.~Aggarwal\Irefn{org86}\And
G.~Aglieri Rinella\Irefn{org36}\And
M.~Agnello\Irefn{org110}\And
N.~Agrawal\Irefn{org47}\And
Z.~Ahammed\Irefn{org130}\And
I.~Ahmed\Irefn{org16}\And
S.U.~Ahn\Irefn{org67}\And
I.~Aimo\Irefn{org93}\textsuperscript{,}\Irefn{org110}
% ...
\renewcommand\labelenumi{\textsuperscript{\theenumi}~}

\section*{Affiliation notes}
\renewcommand\theenumi{\roman{enumi}}
\begin{Authlist}
\item \Adef{0}Deceased
\end{Authlist}

\section*{Collaboration Institutes}
\renewcommand\theenumi{\arabic{enumi}~}
\begin{Authlist}

\item \Idef{org1}A.I. Alikhanyan National Science Laboratory (Yerevan Physics Institute) Foundation, Yerevan, Armenia
\item \Idef{org2}Benem\'{e}rita Universidad Aut\'{o}noma de Puebla, Puebla, Mexico
\item \Idef{org3}Bogolyubov Institute for Theoretical Physics, Kiev, Ukraine
\item \Idef{org4}Bose Institute, Department of Physics and Centre for Astroparticle Physics and Space Science (CAPSS), Kolkata, India
\item \Idef{org5}Budker Institute for Nuclear Physics, Novosibirsk, Russia
\item \Idef{org6}California Polytechnic State University, San Luis Obispo, California, United States
\item \Idef{org7}Central China Normal University, Wuhan, China
\item \Idef{org8}Centre de Calcul de l'IN2P3, Villeurbanne, France
\item \Idef{org9}Centro de Aplicaciones Tecnol\'{o}gicas y Desarrollo Nuclear (CEADEN), Havana, Cuba
\item \Idef{org10}Centro de Investigaciones Energ\'{e}ticas Medioambientales y Tecnol\'{o}gicas (CIEMAT), Madrid, Spain
\item \Idef{org11}Centro de Investigaci\'{o}n y de Estudios Avanzados (CINVESTAV), Mexico City and M\'{e}rida, Mexico
\item \Idef{org12}Centro Fermi - Museo Storico della Fisica e Centro Studi e Ricerche ``Enrico Fermi'', Rome, Italy
\item \Idef{org13}Chicago State University, Chicago, Illinois, USA
\item \Idef{org14}China Institute of Atomic Energy, Beijing, China
\item \Idef{org15}Commissariat \`{a} l'Energie Atomique, IRFU, Saclay, France
\item \Idef{org16}COMSATS Institute of Information Technology (CIIT), Islamabad, Pakistan
\item \Idef{org17}Departamento de F\'{\i}sica de Part\'{\i}culas and IGFAE, Universidad de Santiago de Compostela, Santiago de Compostela, Spain
\item \Idef{org18}Department of Physics and Technology, University of Bergen, Bergen, Norway
\item \Idef{org19}Department of Physics, Aligarh Muslim University, Aligarh, India
\item \Idef{org20}Department of Physics, Ohio State University, Columbus, Ohio, United States
\item \Idef{org21}Department of Physics, Sejong University, Seoul, South Korea
\item \Idef{org22}Department of Physics, University of Oslo, Oslo, Norway
\item \Idef{org23}Dipartimento di Elettrotecnica ed Elettronica del Politecnico, Bari, Italy
\item \Idef{org24}Dipartimento di Fisica dell'Universit\`{a} 'La Sapienza' and Sezione INFN Rome, Italy
\item \Idef{org25}Dipartimento di Fisica dell'Universit\`{a} and Sezione INFN, Cagliari, Italy
\item \Idef{org26}Dipartimento di Fisica dell'Universit\`{a} and Sezione INFN, Trieste, Italy
\item \Idef{org27}Dipartimento di Fisica dell'Universit\`{a} and Sezione INFN, Turin, Italy
\item \Idef{org28}Dipartimento di Fisica e Astronomia dell'Universit\`{a} and Sezione INFN, Bologna, Italy
\item \Idef{org29}Dipartimento di Fisica e Astronomia dell'Universit\`{a} and Sezione INFN, Catania, Italy
\item \Idef{org30}Dipartimento di Fisica e Astronomia dell'Universit\`{a} and Sezione INFN, Padova, Italy
\item \Idef{org31}Dipartimento di Fisica `E.R.~Caianiello' dell'Universit\`{a} and Gruppo Collegato INFN, Salerno, Italy
\item \Idef{org32}Dipartimento di Scienze e Innovazione Tecnologica dell'Universit\`{a} del  Piemonte Orientale and Gruppo Collegato INFN, Alessandria, Italy
\item \Idef{org33}Dipartimento Interateneo di Fisica `M.~Merlin' and Sezione INFN, Bari, Italy
\item \Idef{org34}Division of Experimental High Energy Physics, University of Lund, Lund, Sweden
\item \Idef{org35}Eberhard Karls Universit\"{a}t T\"{u}bingen, T\"{u}bingen, Germany
\item \Idef{org36}European Organization for Nuclear Research (CERN), Geneva, Switzerland
\item \Idef{org37}Faculty of Engineering, Bergen University College, Bergen, Norway
\item \Idef{org38}Faculty of Mathematics, Physics and Informatics, Comenius University, Bratislava, Slovakia
\item \Idef{org39}Faculty of Nuclear Sciences and Physical Engineering, Czech Technical University in Prague, Prague, Czech Republic
\item \Idef{org40}Faculty of Science, P.J.~\v{S}af\'{a}rik University, Ko\v{s}ice, Slovakia
\item \Idef{org41}Faculty of Technology, Buskerud and Vestfold University College, Vestfold, Norway
\item \Idef{org42}Frankfurt Institute for Advanced Studies, Johann Wolfgang Goethe-Universit\"{a}t Frankfurt, Frankfurt, Germany
\item \Idef{org43}Gangneung-Wonju National University, Gangneung, South Korea
\item \Idef{org44}Gauhati University, Department of Physics, Guwahati, India
\item \Idef{org45}Helsinki Institute of Physics (HIP), Helsinki, Finland
\item \Idef{org46}Hiroshima University, Hiroshima, Japan
\item \Idef{org47}Indian Institute of Technology Bombay (IIT), Mumbai, India
\item \Idef{org48}Indian Institute of Technology Indore, Indore (IITI), India
\item \Idef{org49}Inha University, Incheon, South Korea
\item \Idef{org50}Institut de Physique Nucl\'eaire d'Orsay (IPNO), Universit\'e Paris-Sud, CNRS-IN2P3, Orsay, France
\item \Idef{org51}Institut f\"{u}r Informatik, Johann Wolfgang Goethe-Universit\"{a}t Frankfurt, Frankfurt, Germany
\item \Idef{org52}Institut f\"{u}r Kernphysik, Johann Wolfgang Goethe-Universit\"{a}t Frankfurt, Frankfurt, Germany
\item \Idef{org53}Institut f\"{u}r Kernphysik, Westf\"{a}lische Wilhelms-Universit\"{a}t M\"{u}nster, M\"{u}nster, Germany
\item \Idef{org54}Institut Pluridisciplinaire Hubert Curien (IPHC), Universit\'{e} de Strasbourg, CNRS-IN2P3, Strasbourg, France
\item \Idef{org55}Institute for Nuclear Research, Academy of Sciences, Moscow, Russia
\item \Idef{org56}Institute for Subatomic Physics of Utrecht University, Utrecht, Netherlands
\item \Idef{org57}Institute for Theoretical and Experimental Physics, Moscow, Russia
\item \Idef{org58}Institute of Experimental Physics, Slovak Academy of Sciences, Ko\v{s}ice, Slovakia
\item \Idef{org59}Institute of Physics, Academy of Sciences of the Czech Republic, Prague, Czech Republic
\item \Idef{org60}Institute of Physics, Bhubaneswar, India
\item \Idef{org61}Institute of Space Science (ISS), Bucharest, Romania
\item \Idef{org62}Instituto de Ciencias Nucleares, Universidad Nacional Aut\'{o}noma de M\'{e}xico, Mexico City, Mexico
\item \Idef{org63}Instituto de F\'{\i}sica, Universidad Nacional Aut\'{o}noma de M\'{e}xico, Mexico City, Mexico
\item \Idef{org64}iThemba LABS, National Research Foundation, Somerset West, South Africa
\item \Idef{org65}Joint Institute for Nuclear Research (JINR), Dubna, Russia
\item \Idef{org66}Konkuk University, Seoul, South Korea
\item \Idef{org67}Korea Institute of Science and Technology Information, Daejeon, South Korea
\item \Idef{org68}KTO Karatay University, Konya, Turkey
\item \Idef{org69}Laboratoire de Physique Corpusculaire (LPC), Clermont Universit\'{e}, Universit\'{e} Blaise Pascal, CNRS--IN2P3, Clermont-Ferrand, France
\item \Idef{org70}Laboratoire de Physique Subatomique et de Cosmologie, Universit\'{e} Grenoble-Alpes, CNRS-IN2P3, Grenoble, France
\item \Idef{org71}Laboratori Nazionali di Frascati, INFN, Frascati, Italy
\item \Idef{org72}Laboratori Nazionali di Legnaro, INFN, Legnaro, Italy
\item \Idef{org73}Lawrence Berkeley National Laboratory, Berkeley, California, United States
\item \Idef{org74}Lawrence Livermore National Laboratory, Livermore, California, United States
\item \Idef{org75}Moscow Engineering Physics Institute, Moscow, Russia
\item \Idef{org76}National Centre for Nuclear Studies, Warsaw, Poland
\item \Idef{org77}National Institute for Physics and Nuclear Engineering, Bucharest, Romania
\item \Idef{org78}National Institute of Science Education and Research, Bhubaneswar, India
\item \Idef{org79}Niels Bohr Institute, University of Copenhagen, Copenhagen, Denmark
\item \Idef{org80}Nikhef, National Institute for Subatomic Physics, Amsterdam, Netherlands
\item \Idef{org81}Nuclear Physics Group, STFC Daresbury Laboratory, Daresbury, United Kingdom
\item \Idef{org82}Nuclear Physics Institute, Academy of Sciences of the Czech Republic, \v{R}e\v{z} u Prahy, Czech Republic
\item \Idef{org83}Oak Ridge National Laboratory, Oak Ridge, Tennessee, United States
\item \Idef{org84}Petersburg Nuclear Physics Institute, Gatchina, Russia
\item \Idef{org85}Physics Department, Creighton University, Omaha, Nebraska, United States
\item \Idef{org86}Physics Department, Panjab University, Chandigarh, India
\item \Idef{org87}Physics Department, University of Athens, Athens, Greece
\item \Idef{org88}Physics Department, University of Cape Town, Cape Town, South Africa
\item \Idef{org89}Physics Department, University of Jammu, Jammu, India
\item \Idef{org90}Physics Department, University of Rajasthan, Jaipur, India
\item \Idef{org91}Physik Department, Technische Universit\"{a}t M\"{u}nchen, Munich, Germany
\item \Idef{org92}Physikalisches Institut, Ruprecht-Karls-Universit\"{a}t Heidelberg, Heidelberg, Germany
\item \Idef{org93}Politecnico di Torino, Turin, Italy
\item \Idef{org94}Purdue University, West Lafayette, Indiana, United States
\item \Idef{org95}Pusan National University, Pusan, South Korea
\item \Idef{org96}Research Division and ExtreMe Matter Institute EMMI, GSI Helmholtzzentrum f\"ur Schwerionenforschung, Darmstadt, Germany
\item \Idef{org97}Rudjer Bo\v{s}kovi\'{c} Institute, Zagreb, Croatia
\item \Idef{org98}Russian Federal Nuclear Center (VNIIEF), Sarov, Russia
\item \Idef{org99}Russian Research Centre Kurchatov Institute, Moscow, Russia
\item \Idef{org100}Saha Institute of Nuclear Physics, Kolkata, India
\item \Idef{org101}School of Physics and Astronomy, University of Birmingham, Birmingham, United Kingdom
\item \Idef{org102}Secci\'{o}n F\'{\i}sica, Departamento de Ciencias, Pontificia Universidad Cat\'{o}lica del Per\'{u}, Lima, Peru
\item \Idef{org103}Sezione INFN, Bari, Italy
\item \Idef{org104}Sezione INFN, Bologna, Italy
\item \Idef{org105}Sezione INFN, Cagliari, Italy
\item \Idef{org106}Sezione INFN, Catania, Italy
\item \Idef{org107}Sezione INFN, Padova, Italy
\item \Idef{org108}Sezione INFN, Rome, Italy
\item \Idef{org109}Sezione INFN, Trieste, Italy
\item \Idef{org110}Sezione INFN, Turin, Italy
\item \Idef{org111}SSC IHEP of NRC Kurchatov institute, Protvino, Russia
\item \Idef{org112}SUBATECH, Ecole des Mines de Nantes, Universit\'{e} de Nantes, CNRS-IN2P3, Nantes, France
\item \Idef{org113}Suranaree University of Technology, Nakhon Ratchasima, Thailand
\item \Idef{org114}Technical University of Split FESB, Split, Croatia
\item \Idef{org115}The Henryk Niewodniczanski Institute of Nuclear Physics, Polish Academy of Sciences, Cracow, Poland
\item \Idef{org116}The University of Texas at Austin, Physics Department, Austin, Texas, USA
\item \Idef{org117}Universidad Aut\'{o}noma de Sinaloa, Culiac\'{a}n, Mexico
\item \Idef{org118}Universidade de S\~{a}o Paulo (USP), S\~{a}o Paulo, Brazil
\item \Idef{org119}Universidade Estadual de Campinas (UNICAMP), Campinas, Brazil
\item \Idef{org120}University of Houston, Houston, Texas, United States
\item \Idef{org121}University of Jyv\"{a}skyl\"{a}, Jyv\"{a}skyl\"{a}, Finland
\item \Idef{org122}University of Liverpool, Liverpool, United Kingdom
\item \Idef{org123}University of Tennessee, Knoxville, Tennessee, United States
\item \Idef{org124}University of the Witwatersrand, Johannesburg, South Africa
\item \Idef{org125}University of Tokyo, Tokyo, Japan
\item \Idef{org126}University of Tsukuba, Tsukuba, Japan
\item \Idef{org127}University of Zagreb, Zagreb, Croatia
\item \Idef{org128}Universit\'{e} de Lyon, Universit\'{e} Lyon 1, CNRS/IN2P3, IPN-Lyon, Villeurbanne, France
\item \Idef{org129}V.~Fock Institute for Physics, St. Petersburg State University, St. Petersburg, Russia
\item \Idef{org130}Variable Energy Cyclotron Centre, Kolkata, India
\item \Idef{org131}Vin\v{c}a Institute of Nuclear Sciences, Belgrade, Serbia
\item \Idef{org132}Warsaw University of Technology, Warsaw, Poland
\item \Idef{org133}Wayne State University, Detroit, Michigan, United States
\item \Idef{org134}Wigner Research Centre for Physics, Hungarian Academy of Sciences, Budapest, Hungary
\item \Idef{org135}Yale University, New Haven, Connecticut, United States
\item \Idef{org136}Yonsei University, Seoul, South Korea
\item \Idef{org137}Zentrum f\"{u}r Technologietransfer und Telekommunikation (ZTT), Fachhochschule Worms, Worms, Germany
\end{Authlist}
\endgroup